\documentclass[preprint,epsf]{article}
\usepackage{epsf}
\usepackage[T1]{fontenc}
\begin{document}
  \title{Dynamics of Br Electrosorption on Single-Crystal Ag(100):
A Computational Study 
}
\author{
\centerline{S.J. Mitchell, G. Brown, P.A. Rikvold}\\
\centerline{\footnotesize\it Center for Materials Research and Technology,}\\
\centerline{\footnotesize\it Department of Physics, and}\\
\centerline{\footnotesize\it School of Computational Science and 
Information Technology,}\\
\centerline{\footnotesize\it Florida State University, 
Tallahassee, Florida 32306-4351, USA}\\
} 
\date{\today}
\maketitle
\begin{abstract}
We present dynamic Monte Carlo simulations of a lattice-gas model for
bromine electrodeposition on single-crystal silver (100). 
This system undergoes a continuous phase transition between a disordered 
phase at low electrode potentials and a commensurate c($2 \times 2$) 
phase at high potentials. 
The lattice-gas parameters are determined by fitting 
simulated equilibrium adsorption isotherms to chronocoulometric data, 
and free-energy barriers for adsorption/desorption and lateral diffusion are estimated from 
{\it ab initio\/} data in the literature. 
Cyclic voltammograms in the quasi-static limit are obtained by equilibrium
Monte Carlo simulations, while for nonzero potential scan rates
we use dynamic Monte Carlo simulation.
The butterfly shapes of the simulated 
voltammograms are in good agreement with experiments. 
Simulated potential-step experiments give results for the time evolution of 
the Br coverage, as well as the c($2 \times 2$) order parameter and its 
correlation length. 
During phase ordering following a positive potential step, 
the system obeys dynamic scaling.
The disordering following a negative potential step
is well described by random desorption with diffusion.
Both ordering and disordering processes are strongly influenced 
by the ratio of the time scales for desorption and diffusion. 
Our results should be testable by experiments,
in particular cyclic voltammetry and surface X-ray scattering. 
\end{abstract}

\section{Introduction}
\label{sec:I}

The electrodeposition of Br on single-crystal Ag(100) from aqueous
solution is a simple example of anion adsorption which has been
extensively studied, both by classical electrochemical methods 
\cite{VALETTE:AG,OCKO:BR/AG,WANG:FRUM,WAND00,ENDO99}
and by techniques such as {\it in situ\/} surface X-ray scattering (SXS) 
\cite{OCKO:BR/AG,WANG:FRUM,WAND00}
and X-ray absorption fine structure (XAFS) \cite{ENDO99}. 

Cyclic voltammetry (CV) shows a typical butterfly structure with 
a broad pre-wave in the negative-potential region   
and a sharp peak at more positive potentials
\cite{VALETTE:AG,OCKO:BR/AG,WANG:FRUM,WAND00,ENDO99}. 
The recent SXS experiments by Ocko, Wang, and Wandlowski \cite{OCKO:BR/AG} 
showed that the sharp peak corresponds 
to a continuous phase transition in the layer of adsorbed Br.  
At this transition the adlayer changes its structure
from a disordered two-dimensional ``gas'' on the negative-potential
side, to an ordered, commensurate
c($2\times 2$) phase with a Br coverage of 1/2 monolayer on the positive side. 
The pre-wave lies wholly in the disordered-phase region, and 
it was previously suggested that it was 
caused by interactions with surface water \cite{VALETTE:AG}. 
However, equilibrium Monte Carlo (MC)  
simulations of water-free lattice-gas models of the Br adlayer by 
Koper \cite{KOPER:HALIDE,KOPE98C}, as well as our own work reported here,
produce quasi-equilibrium cyclic voltammograms (CVs)
of essentially the same shape as the experiments, indicating that 
surface water is not needed to reproduce the pre-wave. Rather, 
we believe that the pre-wave is
due to short-range correlations in the disordered adlayer
\cite{KOPER:HALIDE,KOPE00}.
In this system, the structures that give rise to 
these correlations {\it locally\/} 
resemble low-temperature ordered phases with coverage 1/4. 

The simplicity of the equilibrium properties of this system makes it a
prime candidate for dynamic studies. Here we present 
results from such a study by
dynamic MC simulation \cite{ME:ECCHAPT} of a lattice-gas model. 
This method has the advantage over mean-field rate-equation approaches that it 
properly accounts for
the effects of local fluctuations in the adlayer structure. Further
details and additional results will be presented elsewhere \cite{MITC00B}. 
Animations and additional figures are available on the World Wide Web
\cite{WWW}.

The rest of this paper is organized as follows.
The lattice-gas model is presented in Sec.~\ref{sec:LGM},
followed by equilibrium simulations which are used to estimate the 
lattice-gas model parameters by fitting to adsorption isotherms from
chronocoulometry experiments \cite{OCKO:BR/AG,WANG:FRUM,WAND00}. 
In Sec.~\ref{sec:dynMC} we present simulated CVs (Sec.~\ref{sec:CV})
and potential-step experiments (Sec.~\ref{sec:step}). 
From the the potential-step simulations we provide current
transients, which are easy to measure in electrochemical experiments, as
well as predictions for the time evolution of the intensity of the 
SXS scattering peak corresponding to the c($2 \times 2$) phase.
Although dynamic SXS data have yet to be obtained for this system,
they were obtained for others \cite{FINN98}, 
and we believe it is only a matter of time before such  
experimental results become available.

\section{Model and Equilibrium Results}
\label{sec:LGM}

We use a lattice-gas model similar to the one used by 
Koper \cite{KOPER:HALIDE}. 
It consists of an $L\times L$ square array of Br adsorption sites, 
corresponding to the four-fold hollow sites on the Ag(100) surface.
The configurational energy of the Br adlayer is given by the 
grand-canonical lattice-gas Hamiltonian,
\begin{equation}
{\cal H}=-\sum_{i<j} \phi_{ij} c_i c_j 
- \overline{\mu} \sum_{i=1}^{L^2} c_i 
\;.
\label{eq:H}
\end{equation}
Here $i$ and $j$ denote adsorption sites,
$c_i$ is the occupation at site $i$,
which is either 0 (empty) or 1 (occupied),
$\sum_{i<j}$ is a sum over all pairs of sites, 
$\phi_{ij}$ is the lateral interaction energy of the pair $(i,j)$,
and $\overline{\mu}$ is the electrochemical potential.
The sign conventions are such 
that $\phi_{ij}<0$ denotes a repulsive interaction,
and $\overline{\mu}>0$ favors adsorption.
We measure $\phi_{ij}$ and 
$\overline{\mu}$ in units of meV/pair and meV/particle, respectively. 
(For brevity, both will be written simply as meV.) 
To reduce finite-size effects, we use periodic boundary conditions.

In the weak-solution 
approximation, $\overline{\mu}$ is related to the electrode potential by 
\begin{equation}
\overline{\mu} = \overline{\mu}_0 
+ k_{\rm B} T \ln \frac{[C]}{[C_{0}]} - e\gamma E 
\;,
\label{eq:mu}
\end{equation}
where $\overline{\mu}_0$ is an arbitrary reference level,
$k_{\rm B}$ is Boltzmann's constant,
$T$ is the absolute temperature,
$e$ is the elementary charge unit,
$[C]$ is the concentration of Br$^-$ in solution, 
$[C_{0}]$ is an arbitrary reference concentration,
$\gamma$ is the electrosorption valency, 
and $E$ is the electrode potential in mV.

Previously Koper has explored the effects of finite nearest-neighbor 
repulsion and screened dipole-dipole interactions in this model 
\cite{KOPER:HALIDE}. As his results indicate only minor effects of the 
finite nearest-neighbor repulsion and screening, we here use a simplified model with
nearest-neighbor exclusion and unscreened dipole-dipole interactions.
Thus, the long-range part of the 
interaction energy is $\phi (r)= { 2^{3/2} \phi_{\rm nnn}}{r^{-3}}$ for
$r \ge \sqrt{2}$, 
where $r$ is the separation of an interacting Br pair, measured in units of 
the Ag(100) lattice spacing ($a=2.889$~{\AA} \cite{OCKO:BR/AG}),
and $\phi_{\rm nnn}$ (which is negative) is the lateral dipole-dipole repulsion 
between next-nearest neighbors. 
{}For computational convenience we cut off the long-range interaction
for $r > 5$. This underestimates the total interaction energy of a fully 
occupied c($2 \times 2$) layer by about $13\%$. 

The nearest-neighbor exclusions give rise to two different sublattices
of possible adsorption sites (like the black and
white squares on a chessboard).
Each sublattice (labeled $A$ and $B$, respectively)
corresponds to one of two degenerate c$(2 \times 2)$ phases.
The $A$ sublattice coverage
is the fraction of occupied sites on sublattice $A$,
defined as $\Theta_A = N_A^{-1} \sum_{i\in A}^{N_A} c_{i}$,
where $N_A$ is the number of sites on sublattice $A$
and $\sum_{i\in A}^{N_A}$ runs over all sites on the sublattice.
We define $\Theta_B$ analogously.

The sublattice coverages combine to give two observables of interest:
the total Br coverage $\Theta = (\Theta_A+\Theta_B)/2$,
and the ``staggered'' coverage $\Theta_{\rm S} = \Theta_A - \Theta_B$,
which is the order parameter for the c$(2 \times 2)$ phase. 
While $\Theta$ can be experimentally obtained by standard electrochemical methods, 
as well as from the integer-order peaks in scattering data, $\Theta_{\rm S}$ 
is proportional to the square root
of the intensity in the half-order diffraction peaks that correspond to the 
c$(2 \times 2)$ phase \cite{OCKO:BR/AG}. 

\begin{figure}
{\epsfysize=3in \epsfbox{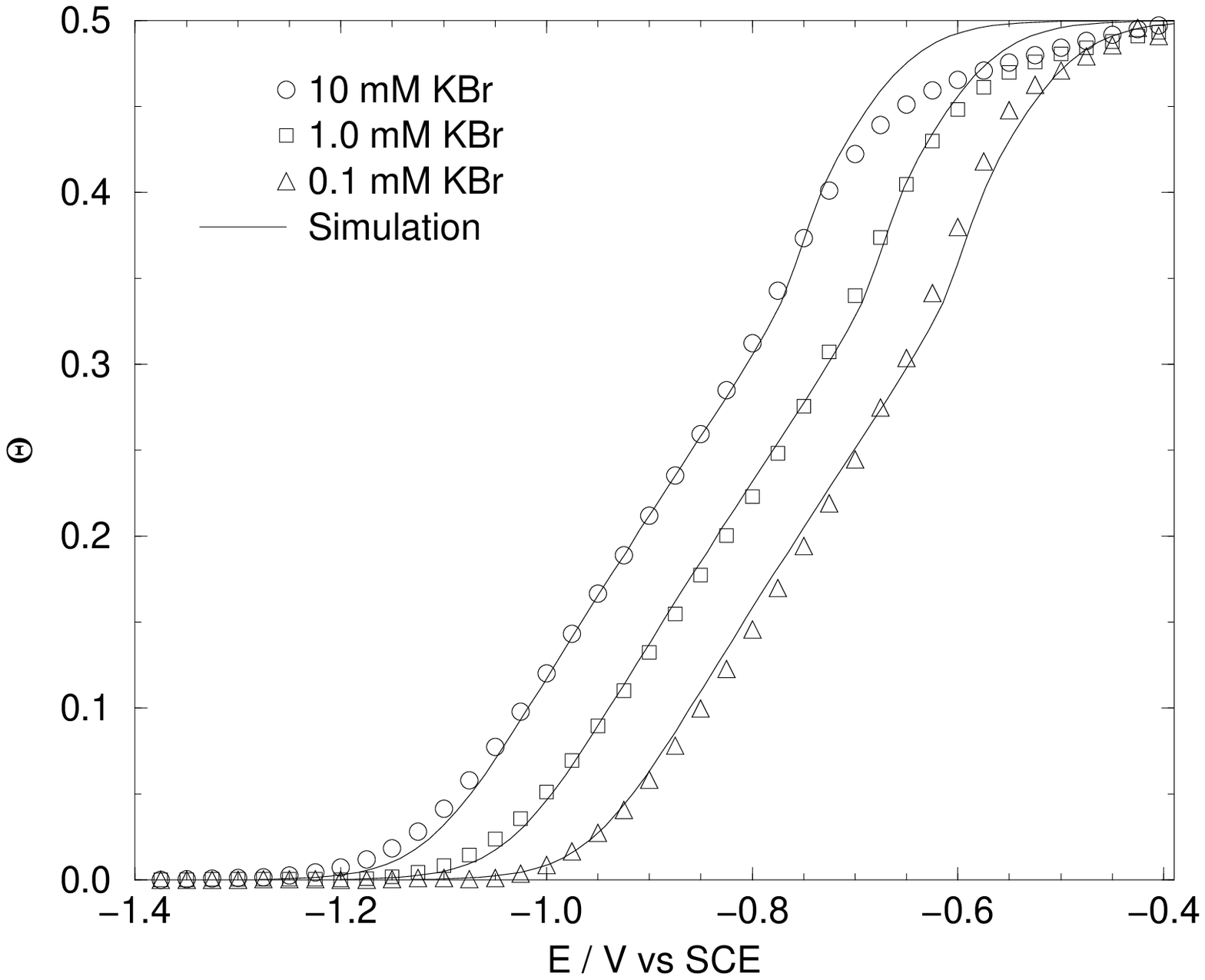}}
\caption[]{
A single equilibrium Monte Carlo (MC) 
simulation isotherm ($L=32$, $T\approx 290$~K), fit simultaneously to 
coverage isotherms at three different Br$^-$ concentrations, 
obtained by chronocoulometry \cite{WANG:FRUM,WAND00}.
The parameters, $\phi_{\rm nnn}=-26\pm2$~meV and $\gamma=-0.73\pm0.03$,
were obtained by a nonlinear fit to the experimental data.
}
\label{fig:exp}
\end{figure}

To estimate the parameters in the lattice-gas model, $\phi_{\rm nnn}$
and $\gamma$, we performed standard equilibrium MC simulations  
\cite{BINDER:MCBOOK} at room temperature ($k_{\rm B} T=25$~meV 
or $T\approx290$~K) to obtain 
$\Theta(\overline{\mu})$ for different parameter values.
These simulated isotherms were then compared with experimental
chronocoulometry data for three different electrolyte concentrations 
\cite{WANG:FRUM,WAND00}, and the best parameter values were determined by a
nonlinear fit \cite{MITC00B}. The resulting values are 
$\phi_{\rm nnn} =-26\pm 2$~meV and $\gamma=-0.73\pm 0.03$.
These results are consistent with those 
found previously \cite{WANG:FRUM,WAND00,KOPER:HALIDE}. 
The fitted MC isotherm is shown together with 
the experimental isotherms in Fig.~\ref{fig:exp}. 
Especially for the two lowest concentrations, 
the agreement is excellent over the whole range of electrode potentials. 
The kinks in the isotherms, observed at $\Theta_c \approx 0.37$ for both
the experiments and simulations, correspond
to the order--disorder phase transition. To within our statistical uncertainty, 
this value of $\Theta_c$ is
consistent with previous results for a square-lattice
model with only nearest-neighbor exclusion, known as the ``hard-square
model'' \cite{REE66,BAXTER:HS,RACZ80}. 

\begin{figure}
{\epsfysize=3in \epsfbox{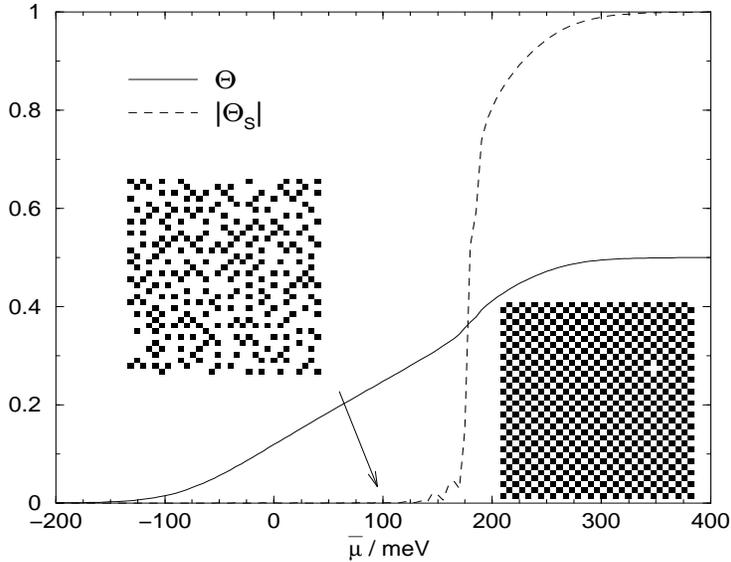}}
\caption[]{
Equilibrium Monte Carlo isotherms for $L=32$, 
$T\approx 290$~K, and $\phi_{\rm nnn}=-26$~meV.
A continuous phase transition 
between a low-coverage disordered phase and a c$(2 \times 2)$
phase with $\Theta=1/2$ occurs at $\bar{\mu}_c \approx 180$~meV.
The order parameter for the c$(2 \times 2)$ phase is $|\Theta_{\rm S}|$. 
These isotherms are obtained from 
10,000 independent samples for each value of $\bar{\mu}$. 
The insets show typical equilibrium configurations in the disordered
phase at $\bar{\mu} = +100$~meV (left) and in the ordered phase at
+400~meV (right). 
}
\label{fig:isoth}
\end{figure}

The simulation data for $\Theta(\bar{\mu})$ and $\Theta_{\rm S}(\bar{\mu})$ 
with the best-fit interaction parameter 
$\phi_{\rm nnn}=-26$~meV are shown together in Fig.~\ref{fig:isoth}. 
As $\bar{\mu}$ approaches its critical value at $\bar{\mu}_c = 180 \pm
5$~meV from the positive side, $\Theta_{\rm S}\propto (\bar{\mu}-\bar{\mu}_c)^{1/8}$. 
This behavior of the order parameter is consistent with the Ising universality class, 
to which this system belongs, and was confirmed by SXS experiments \cite{OCKO:BR/AG}. 
Typical equilibrium configurations are shown as insets in 
Fig.~\ref{fig:isoth} for the disordered phase at 
$\bar{\mu} = +100$~meV and for the ordered phase at $\bar{\mu} = +400$~meV. 

A simulated quasi-equilibrium  CV, corresponding to a vanishing potential-sweep
rate, can be obtained from the MC equilibrium isotherm as  
\begin{equation}
{\cal J}=-\gamma e M \frac{d\Theta}{dt}=
-\gamma e M \frac{d\Theta}{d\bar{\mu}}\frac{d\bar{\mu}}{dE}\frac{dE}{dt}=
\gamma^2 e^2 M\frac{d\Theta}{d\bar{\mu}}\frac{dE}{dt}
\propto \frac{d\Theta}{d\bar{\mu}} \;,
\label{eq:curr}
\end{equation}
where ${\cal J}$ is the voltammetric 
current density (oxidation currents positive), $M$ is the total number of 
adsorption sites per unit area 
($a^{-2} = 1.198 \times 10^{15}$ sites/cm$^2$), 
and ${dE}/{dt}$ is the sweep rate. This limiting CV is shown in  
Figure~\ref{fig:CV}(a) together with CVs for nonzero sweep rates, which
are obtained by dynamic MC simulations discussed in Sec.~\ref{sec:CV}. 
It exhibits the same broad pre-wave and sharp peak seen in experiments 
\cite{VALETTE:AG,OCKO:BR/AG,WANG:FRUM,WAND00,ENDO99}. 
The broad prewave in the simulated CV is caused by
configurational fluctuations in the disordered phase,
which locally resemble low-temperature ordered phases \cite{RIKV91A,RIKV91B}, 
in particular p($2 \times 2$) and c($4 \times 2$) with $\Theta=1/4$ 
\cite{MITC00B}. 
Such fluctuations of local short-range order in the disordered phase are
clearly seen in the inset in Fig.~\ref{fig:isoth} for $\bar{\mu} = +100$meV, 
and they lead to visible anisotropy
in simulated diffuse SXS scattering intensities \cite{MITC00B}. 
This anisotropy should be experimentally observable as well. 

\section{Dynamic Simulations}
\label{sec:dynMC}

The dynamics of the Br adsorption and desorption processes under CV and
potential-step conditions were simulated with a dynamic MC algorithm
including adsorption and desorption events, as well as nearest- and 
next-nearest-neighbor lateral diffusion of the adsorbed Br. 
Bulk diffusion in the solution is neglected, corresponding to a
well-stirred system. Each such single microscopic move, which we 
label by the index $\lambda$, connects an 
initial lattice-gas state, $I$, to a final state, $F$.
The energies of these states, $U_I$ and $U_F$,
are obtained by applying the lattice-gas Hamiltonian, 
Eq.~(\ref{eq:H}), to the corresponding configurations. 
An intermediate transition state of higher energy, $T_\lambda$,  
is associated with the move $\lambda$.
This intermediate state cannot be represented by a lattice-gas configuration,
and we associate with it a ``bare'' free-energy barrier, $\Delta_\lambda$.
Using a symmetric Butler-Volmer approximation
\cite{ME:ECCHAPT,MITC00B,BARD80},
we can then approximate the free energy of the transition state as 
\begin{equation}
U_{T_\lambda}=\frac{U_I+U_F}{2}+\Delta_\lambda \;.
\label{eq:UT}
\end{equation}

Although other choices of the transition probability are also found in the literature
\cite{TAN:TDA}, we here approximate the probability 
$R(F|I)$ of making a transition from $I$ to $F$ during a single MC time step
by the one-step Arrhenius rate \cite{ME:ECCHAPT,KANG89,UEBI97}
\begin{equation}
R(F|I)=\nu \exp{\left(-\frac{U_{T_\lambda}-U_I} {k_{\rm B} T}\right)}
=
\nu \exp{\left(-\frac{\Delta_\lambda}{k_{\rm B} T}\right)}
\exp{\left(-\frac{U_F-U_I}{2 k_{\rm B} T}\right)} \;,
\label{eq:R}
\end{equation}
where the dimensionless rate constant $\nu$ relates the simulation time scale,
measured in MC steps per site (MCSS), 
to the experimental time scale in seconds. In our simulations we used $\nu=1$. 

We performed the simulation with a simple discrete-time dynamical algorithm,
in which new configurations are randomly chosen
from a weighted list of microscopic moves. First, a particular lattice site is 
chosen at random. The configuration can then be changed only through moves which
include the chosen site. 
If the site is empty, adsorption is attempted and accepted with
probability given by Eq.~(\ref{eq:R}). If any of the nearest-neighbor
sites are occupied, this acceptance probability is zero. 
If the chosen site is occupied, only desorption or lateral diffusion may be
attempted. A list is kept of all the possible moves, and one of them is
chosen according to the corresponding acceptance probabilities. The probability 
to remain in the initial state is $R(I|I)=1-\sum_{F \ne I}R(F|I)$.
Further details on the simulation algorithm and its implementation
are given elsewhere \cite{ME:ECCHAPT,MITC00B}. 

While this algorithm is of limited accuracy for very fast processes and
requires an unnecessarily large amount of computer time for very
slow processes \cite{KANG89,BORT75,GILM76,FICH91,NOVO95,LUKK98}, 
it has the advantages that it is easy to program and is readily
adapted to simulations in which parameters (such as $\bar{\mu}$) change
with time \cite{LUKK98}. 
It is thus well suited for dynamic CV simulations. 

The free-energy barriers associated with the different simulation moves are 
$\Delta_{\rm nn}$ for nearest-neighbor diffusion,
$\Delta_{\rm nnn}$ for next-nearest neighbor diffusion, and 
$\Delta_{\rm a}$ for adsorption/desorption. The rough estimates used here, 
$\Delta_{\rm nn}=100$~meV and $\Delta_{\rm nnn}=200$~meV,
are based on {\it ab-initio\/} calculations  
of binding energies for a single Br ion on a Ag(100) substrate in vacuum 
\cite{IGNACZAK:QMHALIDE}.
The difference in binding energy between the bridge site and the four-fold
hollow site gives $\Delta_{\rm nn}$, and 
the difference between the on-top site and the four-fold
hollow site gives $\Delta_{\rm nnn}$.
Theoretical estimates of $\Delta_{\rm a}$ are extremely sensitive to the
ion-surface and ion-water interactions \cite{IGNA98}. 
Calculated potentials of mean force for halide ions in water near a Cu(100) 
surface \cite{IGNA98} indicate that values between 200 and 500~meV are not unreasonable. 
To optimize the simulation speed, we chose $\Delta_{\rm a}=300$~meV.
This value is as low as possible while remaining consistent with our expectation that it 
should be significantly larger than $\Delta_{\rm nnn}$. 

\subsection{Simulated Cyclic Voltammograms}
\label{sec:CV}

\begin{figure}
{\epsfxsize=4in \epsfbox{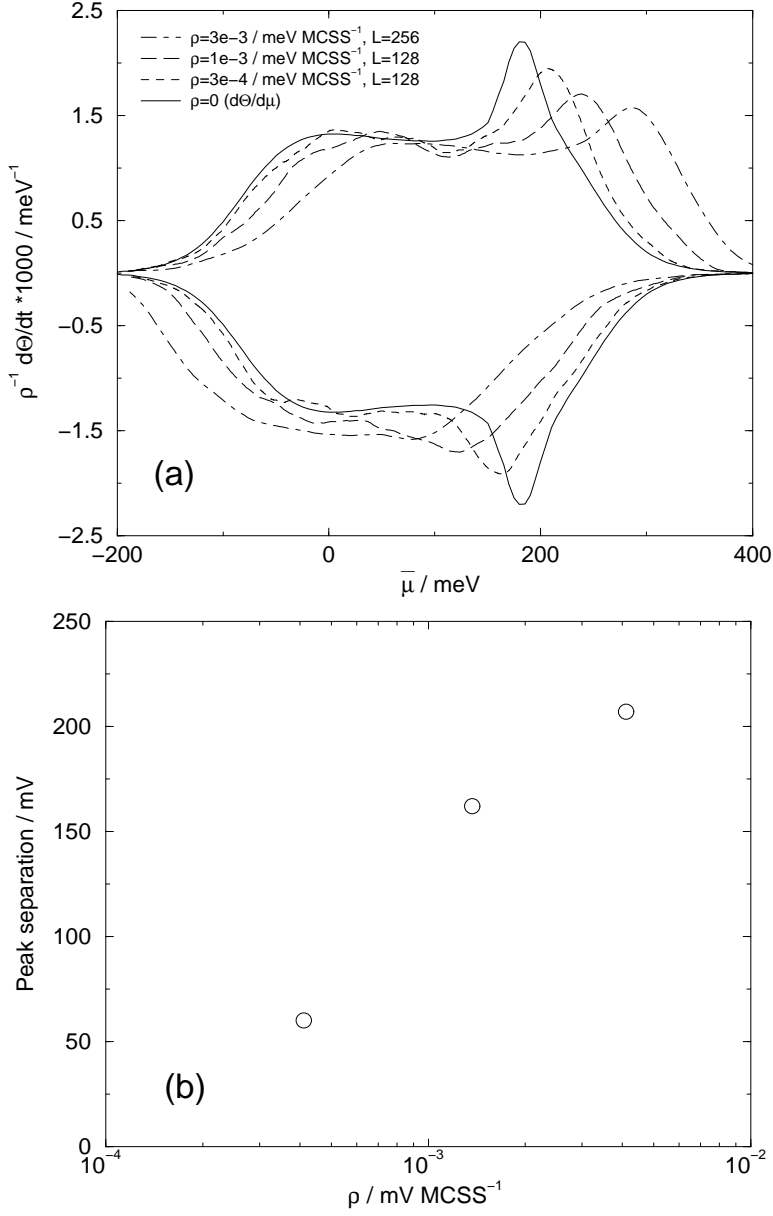}}
\caption[]{
(a) Simulated CVs for $L=128$ and~256 at various potential-sweep rates
$\rho$. The curves for nonzero $\rho$ give $\rho^{-1} d \Theta / d t$,
while for $\rho=0$ we show $d \Theta / d \bar{\mu}$ obtained by
differentiating the simulated equilibrium coverage isotherm in 
Fig.~\protect\ref{fig:isoth}. 
(b) Peak separations for various sweep rates.
For experimental comparison, the units are given as mV and are obtained with $\gamma=-0.73$.
}
\label{fig:CV}
\end{figure}

CV experiments were simulated on systems with $L=256$ and~128 by first 
equilibrating at $\bar{\mu} = -400$~meV and then ramping $\bar{\mu}$ linearly 
in time up to +600~meV, and back down to $-$400~meV. Simulated CV currents 
for sweep rates between 3$\times$10$^{-4}$ and 3$\times$10$^{-3}$~meV/MCSS, 
divided by the sweep rate in order to be easily compared on the same scale, 
are shown in Fig.~\ref{fig:CV}(a). The limiting curve for vanishing sweep rate 
is given by the quasi-equilibrium CV, obtained from the equilibrium simulations
described in Sec.~\ref{sec:LGM}. 
A considerable asymmetry in the peak shape 
for the positive-going and negative-going scans is notable. 

While the pre-waves are relatively little affected by the sweep rate, the peak 
corresponding to the phase transition is significantly lowered and shifted in 
the scan direction. The dependence of the peak separation on the sweep
rate, which is a form of hysteresis \cite{MITC00C}, is shown in 
Fig.~\ref{fig:CV}(b). 

By comparing the scan-rate dependence of the separation between 
the positive-going and negative-going peak positions 
in simulations and experiments, one can in principle obtain a rough estimate of the 
relation between simulation time and physical time, 
provided the free-energy barriers used in the simulations are in a reasonably 
realistic proportion to each other.
The peak separations observed in our simulations, Fig.~\ref{fig:CV}(b), are large compared
to typical experimental values of a few tens of meV
for scan rates in the range 1-10~mV/s.
This most likely indicates that even our slowest simulated scan rates correspond to
faster scans in the real system.
Direct comparisons between simulation and experimental dynamic
effects are left for future research.

\subsection{Simulated Potential Steps}
\label{sec:step}

Dynamic potential-step simulations were performed on systems with $L=256$. 
They began by equilibrating the system
at room temperature and potential $\bar{\mu}_1$. After equilibration, 
$\bar{\mu}$ was instantaneously stepped to a new value, $\bar{\mu}_2$,
and the simulation was continued 
by the dynamic MC algorithm described above.
To illustrate the dynamics of the phase ordering and disordering processes,
we here show results for two different 
potential steps: one from the disordered phase into the ordered phase, 
and one from the ordered phase into the disordered phase. 
Additional potential-step simulations and corresponding time-dependent 
SXS scattering intensities will be reported elsewhere \cite{MITC00B}. 

\subsubsection{Disorder-to-order step}
\label{sec:posstep}

\begin{figure}
{\epsfxsize=4in \epsfbox{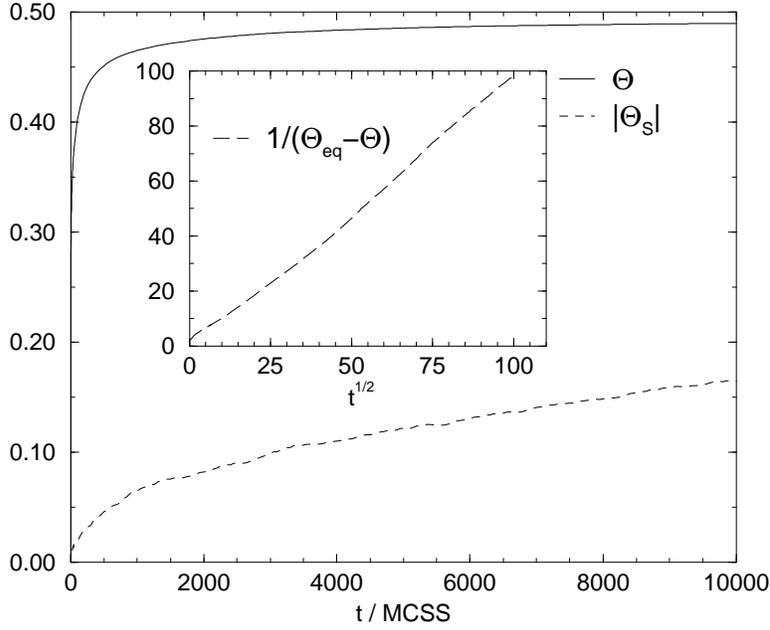}}
\caption[]{
Coverage, $\Theta$, and order parameter, $|\Theta_{\rm S}|$, shown vs time for 
sudden disorder-to-order potential step at room temperature
with $L$=256.
Step from $\bar{\mu}_1=-200$~meV to $\bar{\mu}_2=+600$~meV,
averaged over 10 independent runs.
The inset shows the correlation length for the c$(2\times 2)$ order parameter 
vs $t^{1/2}$.
}
\label{fig:posstep}
\end{figure}

For the disorder-to-order step we used $\bar{\mu}_1=-200$~meV 
and $\bar{\mu}_2=+600$~meV. At $\bar{\mu}_1$ the coverage is close to
zero, while $\bar{\mu}_2$ is far into the ordered phase, 
approximately $420$~meV past the phase transition at
$\bar{\mu}_c$. In Fig.~\ref{fig:posstep} 
we show both $\Theta$ and $\Theta_{\rm S}$ vs time.
With deep steps like this one, the desorption rate is negligible. Thus, the
adsorption dynamics are essentially described by the random sequential
adsorption with diffusion (RSAD) process \cite{WANG93,EISE98}.
The coverage quickly reaches the jamming coverage for
random sequential adsorption of hard squares without diffusion, 
$\Theta_{\rm J} = 0.364$ \cite{MEAK87,BARA95R} 
(only slightly below the critical coverage, $\Theta_c \approx 0.37$). 
At this coverage both sublattices contain small, uncorrelated 
domains of the two degenerate ordered phases, separated by domain walls. 
The domain walls consist of empty sites, most of which cannot be filled, 
due to the nearest-neighbor exclusion. 

At later times, 
the coverage can increase only where domain walls move together, 
opening a gap large enough to fit an additional Br.
The domain-wall motion responsible for opening additional adsorption sites 
proceeds almost exclusively through nearest-neighbor 
diffusion between adjacent domains. The coverage approaches  
the equilibrium value, $\Theta_{\rm eq} = 1/2$, and the total length of 
interfaces decreases. The order-parameter correlation length, $D$, 
is experimentally measurable as the inverse width of the half-order 
diffuse SXS scattering maxima \cite{OCKO:BR/AG,FINN98}. 
Since $D$ is also proportional to the inverse 
of the interfacial length per unit area \cite{DEBY57}, it can in the present case be 
estimated from the coverage as $D = (\Theta_{\rm eq} - \Theta)^{-1}$ 
\cite{WANG93,EISE98}. The dynamic scaling theory of the kinetics of phase 
transitions predicts that, for systems with nonconserved order parameter 
(such as the one considered here) undergoing phase ordering, $D$ 
should grow with 
time as $t^{1/2}$ \cite{GUNTON:DYNTRAN}. The inset in Fig.~\ref{fig:posstep} 
shows $D$ as a function of $t^{1/2}$. 
The agreement with the expected dynamic scaling behavior is excellent, 
as has also been found in previous simulations of RSAD \cite{WANG93,EISE98}. 

{}For an infinite system, $D$ would continue to grow without bound;
however, when $D$ reaches the same order of magnitude as $L$,
the system enters the phase-selection regime in which one of the
two degenerate ordered phases grows to fill the whole system.
Our step simulations were too short to study the phase-selection regime in detail.
Results for shallower positive-going potential steps, which are somewhat less 
clear-cut than for the deep step shown here, will be reported elsewhere 
\cite{MITC00B}. 

\subsubsection{Order-to-disorder step}
\label{sec:negstep}

{}For deep order-to-disorder steps, the disordering process is rather simple. 
Particles desorb at a roughly constant rate,
leading to an essentially exponential relaxation to equilibrium \cite{MITC00B}.

\begin{figure}
{\epsfxsize=4in \epsfbox{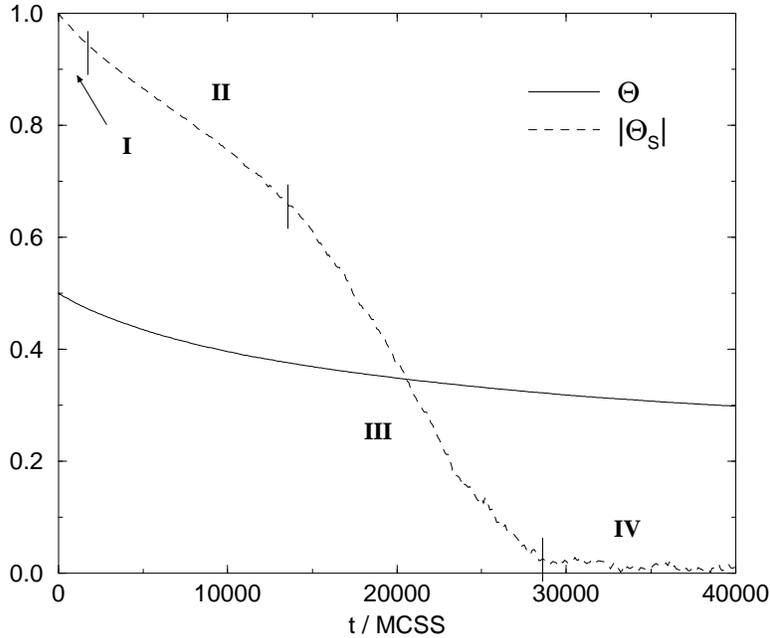}}
\caption[]{
Coverage, $\Theta$, and order parameter, $|\Theta_{\rm S}|$, shown vs time for 
sudden order-to-disorder potential step at room temperature
with $L$=256.
The four dynamic regimes discussed in the text are labeled and separated by vertical bars.
Step from $\bar{\mu}_1=+600$~meV to $\bar{\mu}_2=-200$~meV,
averaged over 3 independent runs. 
}
\label{fig:negstep}
\end{figure}

After a shallow order-to-disorder step, 
$\bar{\mu}_1=+600$~meV and $\bar{\mu}_2=+100$~meV, 
the behavior is more interesting. 
The system starts with all particles on sublattice $A$ and relaxes to a 
disordered phase with $\Theta\approx 1/4$, as shown in Fig.~\ref{fig:negstep}.
There are four different dynamical regimes.
In the first regime, particles simply desorb from sublattice $A$
so that $d\Theta_{\rm S}/dt\approx2\,d\Theta/dt$.
As more sites become vacant, small domains are formed on sublattice $B$
both by lateral diffusion from sublattice $A$ and by adsorption
from the solution. 
In this second regime, both desorption and diffusion contribute significantly
to the disordering process.
In the third regime, the adsorption and desorption 
rates are almost equal and relatively slow, 
and diffusion is the dominant contribution to the disordering.
Here $d\Theta_{\rm S} / d t$ becomes significantly 
larger than $d\Theta/dt$, until the two sublattices become approximately equally populated
in the fourth regime, yielding $\Theta_{\rm S} \approx 0$.

\section{Conclusion}
\label{sec:D}

In this brief paper we have presented results from dynamic MC simulations of 
CV and potential-step experiments for Br electrosorption on Ag(100) 
single-crystal electrodes. The simulations show the complicated interplay 
between adsorption, desorption, and lateral diffusion even in this relatively simple 
electrochemical system. 

The dynamic MC algorithm requires estimates of free-energy barriers for 
adsorption/desorption and lateral diffusion. While such barrier estimates 
constitute the most uncertain part of a dynamic MC simulation, we were able to 
obtain reasonable numbers from {\it ab initio\/} calculations in the 
literature \cite{IGNACZAK:QMHALIDE,IGNA98}.

As the order parameter $\Theta_{\rm S}$ and its correlation length are
measurable in SXS experiments \cite{OCKO:BR/AG,WANG:FRUM,WAND00,FINN98},
the dynamical phenomena predicted by our simulations should be
observable in future experiments.
Further details and simulations of time-dependent
diffuse SXS scattering intensities will be reported elsewhere
\cite{MITC00B}.

\section*{Acknowledgments}
We thank J.~X.\ Wang, T. Wandlowski, and B.~M. Ocko
for providing access to their experimental data
and for useful discussions and correspondence.
We also thank O.~M.\ Magnussen for suggesting the importance of studying 
order-to-disorder potential steps, and V.~Privman for
correspondence on random sequential adsorption.
Supported in part by NSF grants No.\ DMR-9634873 and DMR-9981815,
by Florida State University through
the Center for Materials Research and Technology,
the Supercomputer Computations Research Institute
(U.S.\ Department of Energy contract No. DE-FC05-85FR2500),
and the School of Computational Science and Information Technology.



\clearpage

\end{document}